\documentclass[a4paper,12pt]{article}
\begin{document}

\title {\bf Symmetry induced Dynamics in four-dimensional Models deriving from the 
van der Pol Equation}
\author{Ricardo L\'{o}pez-Ruiz$^{\dagger,\ddagger}$ \\
{\small $^{\dagger}$Department of Computer Science, Faculty of Sciences,} \\
{\small $^{\ddagger}$Instituto de Biocomputaci\'on y F\'isica de Sistemas Complejos,}\\
{\small Universidad de Zaragoza, 50009 - Zaragoza (Spain)}
\date{ }}

\maketitle
\baselineskip 8mm

\begin{center} {\bf Abstract} \end{center}

Different models of self-excited oscillators which are four-dimensional 
extensions of the van der Pol system are reported. 
Their symmetries are analyzed. Three of them were introduced 
to model the release of vortices behind circular cylinders
with a possible transition from a symmetric to an antisymmetric
B\'enard-von Karman vortex street.
The fourth reported self-excited oscillator is a new model which implements
the breaking of the inversion symmetry. It presents the phenomenon 
of second harmonic generation in a natural way.
The parallelism with second harmonic generation in nonlinear 
optics is discussed. There is also a small region in the parameter space  
where the dynamics of this system is quasiperiodic or chaotic.

{\small {\bf PACS number(s)}: 05.45.-a, 47.52.+j}

\newpage

\section{Introduction}

Self-sustained oscillations are found very often 
in mechanical, electrical, biological, chemical and ecological 
systems \cite{andronov}.
Nonlinearities are required in order to attain this kind of behavior.
The system reaches an stable oscillatory dynamics as a consequence of an internal
balance between amplification and dissipation. For instance,
the oscillatory patterns in
some chemical reactions \cite{kuramoto},
the beating of the heart \cite{winfree},  the oscillating
coexistence between different living being species \cite{may}, 
the B\'enard-von Karman vortex street in the wake of a 
cylinder \cite{lopez}, etc. are examples
of continuous and discrete systems modeled as self-excited oscillators.

In this work, five different models of self-excited oscillators
with certain similarities are collected.
First, the fundamental brick of all of them,
the van der Pol equation, is recalled.
Second, the two-dimensional van der Pol equation is extended
to a four-dimensional model in the complex plane. 
Two other models deriving from
the former complex van der Pol system are revisited.
They were investigated as models for the shedding of vortex 
structures in the flow behind a cylinder. 
In a next section, a new model of two coupled self-excited oscillators
implementing the correct symmetries to get second-harmonic
generation is introduced and its qualitative dynamical
behavior is sketched. Finally, the conclusions are presented.

\section{Four-dimensional extensions of the van der Pol Model: 
symmetries and dynamics}

The van der Pol equation, which displays a wide range of behavior
from weakly nonlinear to strongly nonlinear relaxation oscillations,
is one of the simplest models presenting self-oscillations. It reads:
\begin{equation}
\ddot{x}-\epsilon (1-x^2)\dot{x}+x=0,
\label{eq:pol}
\end{equation}
where $x(t)$ is recording the time evolution of a harmonic
oscillator perturbed by the nonlinear dissipative term 
$\epsilon (1-x^2)\dot{x}$. 

The system (\ref{eq:pol}) presents the following symmetries:\newline
\indent (i) {\it Temporal translation symmetry}: 
if $x(t)$ is a solution of equation (\ref{eq:pol})
then $x(t+\phi)$, with $\phi$ a real constant, 
is also a solution of system (\ref{eq:pol}) ; \newline
\indent (ii) {\it Spatial inversion symmetry}:
if $x(t)$ is a solution of the van der Pol equation
then $-x(t)$ verifies also the equation (\ref{eq:pol}). \newline
\indent (iii) {\it $\epsilon$-Parameter inversion symmetry}:
if $x(t)$ is a solution of equation (\ref{eq:pol})
for the constant $\epsilon$ then $-x(-t)$ is also a solution
for the parameter $-\epsilon$. 
Therefore, it is sufficient to find the solutions 
of equation (\ref{eq:pol}) for $\epsilon>0$. 
The dynamics for $\epsilon<0$ is derived by applying 
this symmetry to the latter solutions.

The dynamics of the van der Pol system is quite simple. 
It presents a fixed point at the 
origin and a unique periodic orbit for the whole range of
the parameter $\epsilon$. For $\epsilon<0$, this limit cycle
is unstable, and, for $\epsilon>0$, the limit cycle becomes attractive. 
Due to the uniqueness of this periodic solution, $\bar{x}(t)$,
all the solutions generated from $\bar{x}(t)$ by the symmetries
(i) and (ii) fall on the same orbit in the phase plane representation
given by the coordinates $(x(t),\dot{x}(t))$.
Therefore, there exists a $\bar{\phi}$ satisfying the 
equation $\bar{x}(t+\bar{\phi})=-\bar{x}(t)$,
or equivalently, $\bar{x}(t+2\bar{\phi})=\bar{x}(t)$.
If we call $T$ the period of the limit cycle, that is 
$\bar{x}(t+T)=\bar{x}(t)$, the symmetries of the equation 
(\ref{eq:pol}) impose the condition $2\bar{\phi}=T$.
Then, the final dynamical regime verifies $\bar{x}(t+T/2)=-\bar{x}(t)$.
Let us remark that this property is a consequence of the symmetries 
and the uniqueness of the stable periodic solution of system (\ref{eq:pol}).


The natural extension of the van der Pol equation to 
a four-dimensional model \cite{lopez} can be achieved 
by introducing in the equation (\ref{eq:pol}) the complex variable 
$z=x_1+ix_2$.
Thus we obtain the complex differential equation
\begin{equation}
\ddot{z}-\epsilon (1-|z|^2)\dot{z}+z=0,
\label{eq:pol1}
\end{equation}
where the real variables $(x_1,x_2)$ are the 
coordinates of two symmetric coupled van der Pol self-excited oscillators
which verify: $\ddot{x}_i-\epsilon (1-(x_1^2+x_2^2))\dot{x}_i+x_i=0$,
with $i=1,2$.

Equation (\ref{eq:pol1}) has the symmetries: \newline
\indent (i) {\it Temporal translation symmetry:} 
 $z(t)\rightarrow z(t+\phi)$, with $\phi$ a real constant;\newline
\indent (ii) {\it Phase symmetry:}
 $z(t)\rightarrow e^{i\psi} z(t)$ with $\psi$ real constant.
 It includes the spatial inversion symmetry:
 $z(t)\rightarrow -z(t)$ when $\psi=\pi$;\newline
\indent (iii) {\it Conjugation symmetry:}
 $z(t)\rightarrow\bar{z}(t)$ with $\bar{z}(t)=(x_1(t),-x_2(t))$.\newline
\indent (iv) {\it $\epsilon$-Parameter inversion symmetry:}
 $(z(t),\epsilon)\rightarrow (-z(-t),-\epsilon)$. 

The dynamics of system (\ref{eq:pol1}) 
is damped to zero when $\epsilon<0$ 
and gives self-sustained oscillations when $\epsilon>0$.
These stable periodic motions present the form
$z(t)=e^{i\phi}r(t)$ with $\phi$ being a real constant 
and $r(t)$ satisfying $\ddot{r}-\epsilon (1-r^2)\dot{r}+r=0$.
It can be represented 
as a straight line through the origin in the plane $(x_1,x_2)$
at constant angle $\phi$, $r(t)$ oscillating along this line.
Evidently, this solution verifies $z(t+T/2)=-z(t)$, with $T$ 
the period of the oscillation.
It is found that the trivial periodic solutions $z=e^{\pm it}$ are unstable.


Equation (\ref{eq:pol1}) has been used in Ref. \cite{lopez}
as a fundamental brick to build a simple model 
for the release of vortices behind circular cylinders.
The parameter $\epsilon$ would depend on the Reynolds number
of the flow: $\epsilon<0$ for $Re<Re_c$ and $\epsilon>0$ for 
$Re>Re_c$ where $Re_c$ represents the critical Reynolds number 
for the onset of the vortex shedding \cite{pomeau,benard,vonkarman}. 
By introducing a phase symmetry breaking term controlled 
by the small parameter $\beta$, 
\begin{equation}
\ddot{z}-\epsilon (1-|z|^2)\dot{z}+z+i\beta\bar{z}=0,
\label{eq:pol2}
\end{equation}
the dynamical properties of equation (\ref{eq:pol1}) are modified.
Now there are only two oscillating solutions that asymptotically attract
the flow: the symmetric mode 
$\Theta_1 \equiv (x_1=x_2)$ given by $z_1=e^{\pi/4}r_1(t)$ 
and the antisymmetric mode $\Theta_2\equiv (x_1=-x_2)$
given by $z_2=e^{-\pi/4}r_2(t)$, where $r_1(t)$ and $r_2(t)$ verify
$\ddot{r_i}-\epsilon (1-r_i^2)\dot{r_i}+[1-(-1)^i\beta]r_i =  0$
$(i=1,2)$. Both modes have also the property $z_i(t+T/2)=-z_i(t)$,
$i=1,2$, with $T$ being the period of the motion.
If $\beta>0$, $\Theta_1$ is stable and $\Theta_2$ 
unstable, then the oscillators $(x_1,x_2)$ are in phase 
and the representation of a symmetric vortex 
street is obtained \cite{pomeau}. If $\beta<0$, $\Theta_1$ is unstable 
and $\Theta_2$ stable, then the oscillators $(x_1,x_2)$ are out 
of phase and the representation of an antisymmetric 
vortex street is therefore obtained \cite{benard,vonkarman}.


A slight modification of the nonlinear coupling term in the
dissipative force of equation (\ref{eq:pol2}) produces the model:
\begin{eqnarray}
\ddot{x}_1-\epsilon (1-x_1^2-(1+\gamma )x_2^2))\dot{x_1}+x_1+\beta x_2 & = & 0,\nonumber \\
\ddot{x}_2-\epsilon (1-x_2^2-(1+\gamma )x_1^2))\dot{x_2}+x_2+\beta x_1 & = & 0,
\label{eq:44}
\end{eqnarray} 
where the parameter $\gamma$ controls the breaking of
the conjugation symmetry. The residual symmetries of 
this system in the plane $(x_1,x_2)$ are:\newline
\indent (i) {\it Temporal translation symmetry:} 
 $(x_1(t),x_2(t))\rightarrow (x_1(t+\phi),x_2(t+\phi))$, 
 with $\phi$ a real constant;\newline
\indent (ii) {\it Diagonal reflection symmetry:}
 $(x_1(t),x_2(t))\rightarrow (x_2(t),x_1(t))$; \newline
\indent (iii) {\it $\beta$-Parameter inversion symmetry:}
 $(x_1(t),x_2(t),\beta)\rightarrow (x_2(t),-x_1(t),-\beta)$.
  
A more general and 'robust' unfolding for the transition
between the modes $\Theta_1$ and $\Theta_2$ is obtained
in this new model \cite{lopez}. Two different scenarios 
(Scenario I and Scenario II) for this transition are found.
The nonlinear transition is performed by the mediation of two 
intermediary mixed modes: in Scenario I these mixed modes are 
unstable and in Scenario II they are stable.
The Scenario II has been probably confirmed in the
experiment performed by Zhang {\it et al.} with two flexible filaments
interacting and oscillating in a flowing soap \cite{zhang}.

\section{The Model without Inversion Symmetry}
\label{sec:final}

The inversion symmetry is a fundamental property of the solutions
of the former models. 
We introduce here an inversion symmetry breaking term
in the equation (\ref{eq:pol1}), namely the quadratic term 
$\beta z^2$. We sketch the drastic consequences of this 
term on the dynamical behavior of the system.  
The new equation is
\begin{equation}
\ddot{z}-\epsilon (1-|z|^2)\dot{z}+z+\beta z^2=0,
\label{eq:pol3}
\end{equation}
where $\beta$ is the coupling constant.
The symmetries preserved in this system are: \newline
\indent (i) {\it Temporal translation symmetry}: 
 $z(t)\rightarrow z(t+\phi)$, with $\phi$ a real constant;\newline
\indent (ii) {\it Conjugation symmetry}: $z(t)\rightarrow\bar{z}(t)$;\newline
\indent (iii) {\it $\beta$-Parameter inversion symmetry}:
$(z(t),\beta)\rightarrow (-z(t),-\beta)$. Then it is sufficient to 
study the solutions of equation (\ref{eq:pol3}) for $\beta>0$. The
dynamics for $\beta<0$ is derived by applying this symmetry to the
solutions found for $\beta$ positive.

If we expand the equation (\ref{eq:pol3}) in the real 
variables $(x_1,x_2)$, we have:
\begin{eqnarray}
\ddot{x}_1-\epsilon (1-(x_1^2+x_2^2)) \dot{x}_1+x_1+\beta (x_1^2-x_2^2) & = & 0, \\
\ddot{x}_2-\epsilon (1-(x_1^2+x_2^2)) \dot{x}_2+x_2+2\beta x_1x_2 \;\;\;\;\;\; & = & 0.
\end{eqnarray}
When $\beta$ is different from zero,
a perturbative analysis similar to that performed in Ref. \cite{lopez}
for the model (\ref{eq:44}) does not give in this case any relevant information.
We have performed the numerical integration of these equations  and, roughly,
two different dynamical regimes are found.
The regions where these dynamical behaviors take place
are bounded by two curves, $\beta_1(\epsilon)$ and $\beta_2(\epsilon)$,
in the parameter space $(\epsilon,\beta)$, with $\beta_1<\beta_2$.
For instance, when $\epsilon=1$,  $\beta_1\simeq 0.9$ and $\beta_2\simeq 1.97$,
and the general characteristics of these regions are:\newline
\indent (a) {\it Region I} : if $0<\beta<\beta_1$ or
$\beta>\beta_2$, a unique stable periodic orbit exists in phase space.
It is found that if the coordinate $x_2$ oscillates 
with period $T_2=T$ then the coordinate $x_1$ oscillates 
with the half of a period $T_1=T/2$. The system presents in a natural way
the phenomenon of second-harmonic 
generation in this parameter region (Fig. 1(a)); \newline
\indent (b) {\it Region II} : if $\beta_1<\beta<\beta_2$, the dynamics
of the system displays three different behaviors in the four-dimensional phase space: 
quasiperiodicity (Fig. 1(b)), periodic windows where the relation
$T_2=2T_1$ is also verified (Fig. 1(c)), and chaos (Fig. 1(d)). \newline

The second-harmonic generation phenomenon in Region I
is a consequence of two facts: the symmetries of the system and the existence
of only one stable periodic orbit.
Let $T$ be the period of this orbit, then $(x_1(t+T),x_2(t+T))=(x_1(t),x_2(t))$.
Due to the uniqueness of the stable limit cycle, 
the solutions generated from the time
translation and conjugation transformation of this 
periodic orbit draw the same path on the plane $(x_1,x_2)$. 
There exists therefore a $\bar{\phi}$ verifying that
$(x_1(t+\bar{\phi}),x_2(t+\bar{\phi}))=(x_1(t),-x_2(t))$. 
As in the van der Pol system,
the only coherent solution for the equation $x_2(t+\bar{\phi})=-x_2(t)$ 
is $\bar{\phi}=T/2$. Then, $x_2(t+T/2)=-x_2(t)$ and
$x_1(t+T/2)=x_1(t)$, that is $T_2=2T_1=T$.
The oscillator $x_1$ performs two oscillations for each oscillation
of the oscillator $x_2$. The limit cycle on the plane $(x_1,x_2)$
presents hence a pair number of lobes in the direction of the 
$x_2$ axis and it shows an eight-folded shape (Fig. 1(a-c)). 

Let $z(t,\beta)$ be the unique stable periodic orbit in the Region I.
By numerical inspection
we find out that $z(t,\beta=0)=z_0(t)=(0,x_{2,0}(t))$,
with $x_{2,0}(t)$ verifying 
$\ddot{x}_{2,0}-\epsilon (1-x_{2,0}^2)\dot{x}_{2,0}+x_{2,0}=0$.
In order to perform an alternative perturbation analysis \cite{hinch}
in the parameter $\beta$, we expand $z(t,\beta)=(x_1(t,\beta),x_2(t,\beta))$ 
in power series of $\beta$:
\begin{eqnarray}
x_1(t,\beta) & = & \sum_{j=1}^{\infty}\beta^jx_{1,j}(t), \\
x_2(t,\beta) & = & x_{2,0}(t)+\sum_{j=1}^{\infty}\beta^jx_{2,j}(t),
\end{eqnarray}
where $x_{i,j}(t)$ ($i=1,2$ ; $j=1,\ldots,\infty$) is calculated
by an iterative process as a function of $x_{2,0}(t)$.
Then, $x_{i,j}(t)=\tilde{x}_{i,j}(x_{2,0}(t))$ ,
with $i=1,2$ and $j=1,\ldots,\infty$. The tilde of $\tilde{x}_{i,j}$ indicates 
the different functional dependence of $x_{i,j}$ on $x_{2,0}(t)$. 
Thus, we have:
\begin{eqnarray}
\tilde{x}_1(x_{2,0}(t),\beta) & = & 
              \sum_{j=1}^{\infty}\beta^j\tilde{x}_{1,j}(x_{2,0}(t)), \\
\tilde{x}_2(x_{2,0}(t),\beta) & = & 
              x_{2,0}(t)+\sum_{j=1}^{\infty}\beta^j\tilde{x}_{2,j}(x_{2,0}(t)).
\end{eqnarray} 
  
The $\beta$-parameter inversion symmetry 
$(z(t),\beta)\rightarrow (-z(t),-\beta)$ implies that the relations:
\begin{eqnarray}
\tilde{x}_1(-x_{2,0}(t),-\beta) & = & -\tilde{x}_1(x_{2,0}(t),\beta),\label{eq:1} \\
\tilde{x}_2(-x_{2,0}(t),-\beta) & = & -\tilde{x}_2(x_{2,0}(t),\beta),
\end{eqnarray} 
must be verified by the correct power series expansion.
Furthermore, the conjugation symmetry 
$(z(t),\beta)\rightarrow (\bar{z}(t),\beta)$
imposes the conditions: 
\begin{eqnarray}
\tilde{x}_1(-x_{2,0}(t),\beta) & = & \tilde{x}_1(x_{2,0}(t),\beta), \\
\tilde{x}_2(-x_{2,0}(t),\beta) & = & -\tilde{x}_2(x_{2,0}(t),\beta).
\label{eq:2}
\end{eqnarray} 
Retaining the correct terms which verify the parity conditions
(\ref{eq:1}-\ref{eq:2}), we arrive at the correct power series expansion:  
\begin{eqnarray}
\tilde{x}_1 & = & \sum_{j=1}^{\infty}\beta^{2j-1}\tilde{x}_{1,2j-1}, \label{eq:3}\\
\tilde{x}_2 & = & x_{2,0}+\sum_{j=1}^{\infty}\beta^{2j}\tilde{x}_{2,2j},\label{eq:4}
\end{eqnarray}
where $\tilde{x}_{2,2j}$ are odd functions of $x_{20}$ and
$\tilde{x}_{1,2j-1}$ are even functions of $x_{20}$ for all $j$.
This last property implies
that $x_1$ is a function of $x_{2,0}^2$. 
Because $x_{2,0}(t+T/2)=-x_{2,0}(t)$ is verified
we find that $x_{2,0}^2(t+T/2)=x_{2,0}^2(t)$.
Therefore the frequency associated to the oscillator $x_1(t)$
is the doubled frequency of the oscillator $x_2(t)$, as 
it has been shown above by the argument of the symmetries.

As an example, the first terms of the expansion (\ref{eq:3}-\ref{eq:4}),
$x_1\simeq \beta x_{1,1}$
and $x_2\simeq x_{2,0}+\beta^2 x_{2,2}$, satisfy the differential equations:
\begin{eqnarray}
\ddot{x}_{2,0}-\epsilon (1-x_{2,0}^2)\dot{x}_{2,0}+x_{2,0} & = & 0 \label{eq:1*}, \\
\ddot{x}_{1,1}-\epsilon (1-x_{2,0}^2)\dot{x}_{1,1}+x_{1,1} & = & x_{2,0}^2, \label{eq:2*}\\
\ddot{x}_{2,2}-\epsilon (1-x_{2,0}^2)\dot{x}_{2,2}+[1+\epsilon\dot{(x_{2,0}^2)}]x_{2,2}
& = & -2x_{2,0}x_{1,1}-\epsilon\dot{x}_{2,0}x_{1,1}^2 \label{eq:3*}.
\end{eqnarray}
We have verified the correctness of this expansion numerically,
by recording $x_{2,0}(t)$ from equation (\ref{eq:1*}) and by reinjecting
this signal in equations (\ref{eq:2*}-\ref{eq:3*}), thus obtaining
$x_{1,1}(t)$ and $x_{2,2}(t)$. The comparison of this approximation
with the real dynamics of $x_1(t)$ and $x_2(t)$ derived from 
equation (\ref{eq:pol3}) presents a good agreement
for $\beta$ slightly different from zero. Also,  
the dependence on the parameter $\beta$ of the limit cycle period 
has been calculated by computational methods for different values 
of $\epsilon$. The results are shown in Fig. 2.

The region of the parameter space 
where the system is chaotic is also investigated by computational methods.
We apply directly the method of calculating the sensitive dependence
on initial conditions for each particular pair $(\epsilon,\beta)$
of parameter values. When the system shows exponential dependence on 
the initial conditions, that is, when two infinitesimal separated 
initial conditions become separated of the system size order, 
then the system is chaotic and the pair of parameters is collected. 
The location of the chaotic region in the 
parameter space $(\epsilon,\beta)$ is shown in Fig. 3.\par

Second-harmonic generation is one the 
first phenomena studied in nonlinear optics \cite{svelto}.
Its discovery \cite{franken} put in evidence the possibility
to modify the optical properties of a material system under the action
of an applied optical field sufficiently intense (laser). Thus, in that process,
a fraction of the power in the incident radiation at frequency $w$ is converted
to radiation at the second-harmonic frequency $2w$ (Fig. 4). 
One of the causes of this phenomenon is the non-vanishing 
second-order susceptibility $\chi^{(2)}$ 
of the crystal on which the laser beam is incident. 
This possibility is only achieved in crystals which do not display
inversion symmetry. 
Therefore one would be tempted to establish
some formal relation between the behavior of the model (\ref{eq:pol3})
in the Region I and that optical system (Fig. 4).
From a qualitative point of view, this model presents the breaking
of the inversion symmetry and the frequency doubling phenomenon 
in a natural way.

\section{Conclusions}

In the past few years, important geometrical and topological studies 
have been developed for understanding the periodic orbit organization 
of low dimensional chaotic attractors \cite{mindlin,gilmore}.   
However, we still experience, for instance, 
some difficulty to get a good visual understanding 
of the entangled dynamics of a particle
following a chaotic trajectory in a strange attractor,
even if the system is a three dimensional one.
Probably, this a consequence of the lack of some future technical 
development that would improve and would approach better 
this visual representation more than a consequence of the 
lack of theoretical tools for its comprehension.

Thus, the study of four-dimensional systems can be a challenge
in the next future. In this work, different self-excited 
oscillators which are four-dimensional extensions
of the van der Pol equation have been reported.
Symmetries play a determinant role in the 
dynamics of all of them. In the model (\ref{eq:pol1}),
the circular symmetry forbids the capability 
of this system to select a final dynamical state 
with a fixed symmetry. By introducing a  
phase-symmetry breaking term, $i\beta\bar{z}$, 
and depending of the sign of the control parameter $\beta$,
the dynamics in the model (\ref{eq:pol2}) can settle down on two 
different oscillation states: 
an in-phase mode $\Theta_1$ and another out-of-phase mode 
$\Theta_2$. Each mode might represent a symmetric and an antisymmetric 
B\'enard-von Karman vortex street, respectively \cite{lopez}.  
In the model (\ref{eq:44}), an additional breaking circular symmetry 
term in the dissipative force, monitored by the parameter $\gamma$, 
provokes the existence of two possible scenarios 
for the nonlinear transition between 
the two oscillation modes, $\Theta_1$ and $\Theta_2$.

Let us observe that the final model (\ref{eq:pol3}) implements the 
necessary breaking term of the inversion symmetry, $\beta z^2$,
in order to obtain second-harmonic generation when 
two identical self-excited oscillators interact.
When $\beta=0$ the self-excited oscillators are uncoupled and 
they oscillate to the same frequency $w$. 
It can represent a laser beam before reaching 
a crystal with non-vanishing 
second-order susceptibility. When laser beam
passes through the crystal a nonlinear interaction 
takes place with the medium.
Then the two self-excited oscillators are coupled ($\beta\neq 0$). 
One of them preserves the
original frequency $w$ and the other one oscillates to
the doubled frequency $2w$ (Fig. 4).
Other more complicated phenomena such as quasiperiodicity and chaos 
are also found in a small and very localized region of the parameter space
of the model (\ref{eq:pol3}) (Region II ). 
In the case of its experimental existence,
the fact that the two signals become chaotic in this regime
as a consequence of the nonlinear coupling caused by the crystal would
convert this unseen behavior in a surprising phenomenon.

\newpage
\begin{center} {\bf Figure Captions} \end{center}

{\bf Fig 1.} Projections on the plane  $(x_1,x_2)$ of different 
stable attractors of system (\ref{eq:pol3}).   
{\bf (a)} Projections of the stable limit cycle 
for $\epsilon=1$ and $\beta=0.15$. 
The biggest orbit represents the projection on
the plane $(x_2,\dot{x}_2)$. The smallest one is the projection on 
the plane $(x_1,\dot{x}_1)$. The other eight-shaped orbit is the projection 
on the plane $(x_1,x_2)$. 
{\bf (b)} Projection of the quasiperiodic attractor
for $\epsilon=1$ and $\beta=1$. 
{\bf (c)} Projection of a stable orbit in a periodic window
for $\epsilon=1$ and $\beta=1.2$. Also, in this case, $T_2=2T_1$. 
{\bf (d)} Projection of the chaotic attractor for 
$\epsilon=1$ and $\beta=1.6$. 

{\bf Fig 2.} {\bf (a)} Plot of the normalised period, $T/2\pi$, of the 
limit cycle of the van der Pol oscillator as a function of the 
nonlinear friction strength given by the parameter $\epsilon$. 
{\bf (b)} Plot of the normalised period, $T/2\pi$, of the limit cycle of
system (\ref{eq:pol3}) as a function of the coupling parameter $\beta$ 
for $\epsilon=0.5$ (triangles), $\epsilon=1$ (circles), 
$\epsilon=1.5$ (diamonds) and $\epsilon=2$ (squares).

{\bf Fig 3.} Values of the parameters $(\epsilon,\beta)$ where the system 
(\ref{eq:pol3}) is chaotic. The sensitive dependence on initial conditions 
has been verified by computational methods. 

{\bf Fig 4.} Schematic representation of the second-harmonic generation
phenomenon in nonlinear optics.

\end{document}